\documentclass[12pt]{article}
\usepackage{epsf,epsfig}

\newcommand{\autor}[1] {\begin{center}{\bf \lineskip .3cm #1} \end{center}}
\newcommand{\address}[1] {\begin{center}  {\normalsize \bf \it #1 }\end{center}}
\def\simgt{\rlap{\lower 3.5 pt \hbox{$\mathchar \sim$}} \raise 1pt \hbox {$>$}}
\def\simlt{\rlap{\lower 3.5 pt \hbox{$\mathchar \sim$}} \raise 1pt \hbox {$<$}}

\def\3half{\textstyle\frac32}
\begin{document}
\begin{titlepage}
\vspace{1.0cm}
\begin{center}
\large\bf\boldmath $B_c(B){\to}D l\tilde{\nu}$ form factors in
Light-Cone Sum Rules and the $D$-meson distribution amplitude
\unboldmath
\end{center}
\vspace{0.8cm} \autor{{Fen Zuo\footnote{Email:
zuof@mail.ihep.ac.cn}} and {Tao Huang\footnote{Email:
huangtao@mail.ihep.ac.cn}}} \vspace{0.7cm}
\address{ Institute of High Energy Physics, P.O.Box 918 ,
Beijing 100049, P.R.China\,\footnote{Mailing address}}
\vspace{1.0cm}
\begin{abstract}
In this paper we calculate the weak form factors of the decays
$B_c(B)\to Dl\tilde\nu$ by using the chiral current correlator
within the framework of the QCD light-cone sum rules (LCSR). The
expressions of the form factors only depend on the leading twist
distribution amplitude (DA) of the $D$ meson. Three models of the
$D$-meson distribution amplitude are employed and the calculated
form factor $F_{B_c\to D}(0)$ is given. Our prediction, by using the
$D$-meson distribution amplitude with the exponential suppression at
the end points, is compatible with other approaches, and favors the
three-points sum rules (3PSR) approach with the Coulumb corrections
included.\\
\\
PACS number(s): 13.20.He, 13.20.Fc, 11.55.Hx

\end{abstract}

\end{titlepage}

\section{Introduction}
~~~The CDF Collaboration reported on the observation of the
bottom-charm $B_c$ meson at Fermilab \cite{CDF1} in the semileptonic
decay mode $B_c\to J/\psi+l+\nu$ with the $J/\psi$ decaying into
muon pairs in $1998$. Values for the mass and the lifetime of the
$B_c$ meson were given as $M(B_c)=6.40\pm0.39\pm0.13$ GeV and
$\tau(B_c)=0.46^{+0.18}_{-0.16}(\mbox{stat})\pm0.03(\mbox{syst})$
ps. Recently, CDF reported first Run {I}{I} evidence for the $B_c$
meson in the fully reconstructed decay channel $B_c\to J/\psi+\pi$
with $J/\psi\to\mu^+\mu^-$ \cite{CDF2}. The mass value quoted for
this decay channel is
$6.2857\pm0.0053(\mbox{stat})\pm0.0012(\mbox{syst})$ GeV with errors
significantly smaller than in the first measurement. Also D0 has
observed the $B_c$ in the semileptonic mode $B_c\to J/\psi+\mu+X$
and reported preliminary evidence that
$M(B_c)=5.95^{+0.14}_{-0.13}\pm0.34$ GeV and
$\tau(B_c)=0.45^{+0.12}_{-0.10}\pm0.12$ ps \cite{D0}.

The $B_c$ decays, at first, calculated in the potential models (PM)
\cite{PM1,PM2}, wherein the variation of techniques results in close
estimates after the adjustment on the semileptonic decays of $B$
mesons. The Operator Product Expansion (OPE) evaluation of inclusive
decays gave the lifetime and widths \cite{OPE}, which agree well
with PM, if one sums up the dominating exclusive modes. That was
quite unexpected, when the sum rules (SR) of QCD results in the
semileptonic $B_c$ widths \cite{3PSR1}, which are one order of
magnitude less than those of PM and OPE. The reason may be the
valuable role of Coulomb corrections, that implies the summation of
$\alpha_s/v$ corrections significant in the heavy quarkonia, i.e. in
the $B_c$ \cite{Coulomb}.

   In the recent paper \cite{BD}, we calculate the form factor for
$B{\to}D l\tilde{\nu}$ transitions within the framework of  QCD
light-cone sum rules (LCSR). In the velocity transfer region $1.14 <
v\cdot v' < 1.59$, which renders the OPE near light-cone $x^2=0$ go
effectively, the yielding behavior of form factor is in agreement
with that extracted from the data on $B\to D l\tilde{\nu}$, within
the error. In the larger recoil region $1.35 < v\cdot v' < 1.59$,
the results are observed consistent with those of perturbative QCD
(pQCD).  In this paper we calculate the form factor of the
semileptonic decay $B_c{\to}D l\tilde{\nu}$, which also depends on
the $D$-meson DA. However, due to the different feature of the two
process, the $c$ quark is a spectator in the decay $B_c{\to}D
l\tilde{\nu}$ and the $c$ quark comes from the $b$ quark decay in
the process $B\to D l\tilde{\nu}$, these two form factors are
sensitive to the shape of the DA in two different regions. Combining
the information in the two process, we can find which model is more
suitable for describing the $D$ meson. Similar to the case of $B \to
\pi l\tilde{\nu}$, the LCSR approach for the $B_c{\to}D
l\tilde{\nu}$ form factor is reliable only in the region
$0<q^2<15\mbox{GeV}^2$. we extrapolate the result to the whole
region and give the decay width and branching ratio for the
semileptonic decay.

 This paper is organized as follows. In the following section we
derive the LCSRs for the form factor of $B_c\to Dl\tilde\nu$. A
discussion of the DA models for the $D$ meson is given in section 3.
Section 4 is devoted to the numerical analysis and comparison with
other approaches. The last section is reserved for summary.

\section{LCSRs for the $B_c(B){\to}D$ Form Factors}

~~~The $B_c\to D$ weak form factors $f(q^2)$ and $\tilde{f}(q^2)$
are usually defined as:
\begin{equation}
{<}D(p)|\bar{u}\gamma_{\mu}b|B_c(p+q){>}=2f(q^2)p_\mu+\tilde{f}(q^2)q_\mu,\label{eq:def}
\end{equation}
with $q$ being the momentum transfer.

To achieve a LCSR estimate of $f(q^2)$, we follow Ref.\cite{cLCSR}
and use the following chiral current correlator $\Pi_\mu(p,q)$:
\begin{eqnarray}
\Pi_\mu(p,q)&=&i\int{d^4xe^{ipx}{<}D(p)|T\{\bar{u}(x)\gamma_\mu(1+\gamma_5)b(x),
\bar{b}(0)i(1+\gamma_5)c(0)\}|0{>}}\nonumber\\
&=&\Pi(q^2,(p+q)^2)p_\mu+\tilde{\Pi}(q^2,(p+q)^2)q_\mu,\label{eq:cc}
\end{eqnarray}

First, we express the hadronic representation for the correlator.
This can be done by inserting the complete intermediate states with
the same quantum numbers as the current operator
$\bar{b}i(1+\gamma_5)c$. Isolating the pole contribution due to the
lowest pseudoscalar $B_c$ meson, we have the hadronic representation
in the following:
\begin{eqnarray}
\Pi_{\mu}^H(p,q)&=&\Pi^H(q^2,(p+q)^2)p_\mu+\tilde{\Pi}^H(q^2,(p+q)^2)q_{\mu}\nonumber\\
&=&\frac{{<}D|\bar{u}\gamma_{\mu}b|B_c{>}{<}B_c|\bar{b}i\gamma_5c|0{>}}{m^2_{B_c}-(p+q)^2}\nonumber\\
&&+\sum_H\frac{{<}D|\bar{u}\gamma_{\mu}(1+\gamma_5)b|B_c^H{>}{<}B_c^H|\bar{b}i(1+\gamma_5)c|0{>}}{m^2_{B_c^H}-(p+q)^2}.
\end{eqnarray}
Note that the intermediate states $B_c^H$ contain not only the
pseudoscalar resonance of masses greater than $m_{B_c}$, but also
the scalar resonances with $J^P=0^+$, corresponding to the operator
$\bar{b}c$. With Eq.(\ref{eq:def}) and the definition of the decay
constant $f_{B_c}$ of the $B_c$ meson
\begin{equation}
{<}B_c|\bar{b}i\gamma_5c|0{>}=m_{B_c}^2f_{B_c}/(m_b+m_c),
\end{equation}
and expressing the contributions of higher resonances and continuum
states in a form of dispersion integration, the invariant amplitudes
$\Pi^H$ and $\tilde\Pi^H$ read,
\begin{equation}
\Pi^H[q^2,(p+q)^2]=\frac{2f(q^2)m_{B_c}^2f_{B_c}}{(m_b+m_c)(m_{B_c}^2-(p+q)^2)}+
\int^\infty_{s_0}{\frac{\rho^H(s)}{s-(p+q)^2}ds}+\mbox{subtractions},
\end{equation}
and
\begin{equation}
\tilde{\Pi}^H[q^2,(p+q)^2]=\frac{\tilde{f}(q^2)m_{B_c}^2f_{B_c}}{(m_b+m_c)(m_{B_c}^2-(p+q)^2)}+
\int^\infty_{s_0}{\frac{\tilde{\rho}^H(s)}{s-(p+q)^2}ds}+\mbox{subtractions},
\end{equation}
where the threshold parameter $s_0$ should be set near the squared
mass of the lowest scalar $B_c$ meson, the spectral densities
$\rho^H(s)$ and $\tilde{\rho}^H(s)$ can be approximated by invoking
the quark-hadron duality ansatz
\begin{equation}
\rho^H(s)(\tilde\rho^H(s))=\rho^{QCD}(s)(\tilde\rho^{QCD}(s))\theta(s-s_0).
\end{equation}

On the other hand, we need to calculate the correlator in QCD theory
to obtain the desired sum rule result. In fact, there is an
effective kinematical region which makes OPE applicable:
$(p+q)^2-m_b^2{\ll}0$ for the $b\bar{d}$ channel and
$q^2{\le}m_b^2-2\Lambda_{QCD}m_b$ for the momentum transfer. For the
present purpose, it is sufficient to consider the invariant
amplitude $\Pi(q^2,(p+q)^2)$ which contains the desired form factor.
The leading contribution is derived easily by contracting the
$b-$quark operators to a free propagator:
\begin{equation}
{<0}|Tb(x)\bar{b}(0)|{0>}=\int{\frac{d^4k}{{(2\pi)}^4}e^{-ikx}\frac{k\!\!\!/+m_b}{k^2-m_b^2}}.\label{eq:pro}
\end{equation}\
Substituting Eq.(\ref{eq:pro}) into Eq.(\ref{eq:cc}), we have the
two-particle contribution to the correlator,
\begin{equation}
\Pi_\mu^{(\bar{q}q)}=-2m_bi\int{\frac{d^4xd^4k}{(2\pi)^4}e^{i(q-k)x}\frac{1}{k^2-m_b^2}{<}D(p)|T\bar{c}(x)\gamma_{\mu}\gamma_5d(0)|{0>}}.\label{eq:qq}
\end{equation}
An important observation, as in Ref.\cite{cLCSR}, is that only the
leading non-local matrix element
${<}D(p)|\bar{u}(x)\gamma_{\mu}\gamma_5c(0)|0{>}$ contributions to
the correlator, while the nonlocal matrix elements
${<}D(p)|\bar{u}(x)i\gamma_5c(0)|0{>}$ and
${<}D(p)|\bar{u}(x)\sigma_{\mu\nu}\gamma_5c(0)|0{>}$ whose leading
terms are of twist $3$, disappear from the sum rule. Proceeding to
Eq.(\ref{eq:qq}), we can expand the nonlocal matrix element
${<}D(p)|T\bar{u}(x)\gamma_{\mu}\gamma_5c(0)|0{>}$ as
\begin{equation}
{<}D(p)|T\bar{u}(x)\gamma_{\mu}\gamma_5c(0)|0{>}=-ip_{\mu}f_D\int_0^1{due^{iupx}\varphi_D(\bar{u})}+\mbox{higher
twist terms},\label{eq:da}
\end{equation}
where $\varphi_D(\bar{u})$ is the twist-2 DA of the $D$ meson with
$\bar{u}=1-u$ being the longitudinal momentum fraction carried by
the $c$ quark, those DA's entering the higher-twist terms are of at
least twist $4$. The use of Eq.(\ref{eq:da}) yields
\begin{equation}
\Pi^{(\bar{q}q)}[q^2,(p+q)^2]=2f_Dm_b\int_0^1{du\frac{\varphi_D(\bar{u})}{m_b^2-(up+q)^2}}+\mbox{higher
twist terms}. \label{eq:qq1}
\end{equation}

Invoking a correction term due to the interaction of the $b$ quark
with a background field gluon into Eq.(\ref{eq:qq1}), the
three-particle contribution $\Pi^{(\bar{q}qg)}_\mu$ is achievable.
However, the practical calculation shows that the corresponding
matrix element whose leading term is of twist $3$ also vanishes.
Thus, if we work to the twist-3 accuracy, only the leading twist DA
$\varphi_D$ is needed to yield a LCSR prediction.

Furthermore, we carry out the subtraction procedure of the continuum
spectrum, make the Borel transformations with respect to $(p+q)^2$
in the hadronic and the QCD expressions, and then equate them.
Finally, we get the LCSR for $f(q^2)$:
\begin{equation}
f_{B_c\to
D}(q^2)=\frac{m_b(m_b+m_c)f_D}{m_{B_c}^2f_{B_c}}e^{m_{B_c}^2/M^2}\int_{\Delta_{B_c}}^1{\frac{du}{u}\exp{\left[-\frac{m_b^2-(1-u)(q^2-um_D^2)}{uM^2}\right]}
\varphi_D(\bar{u})},\label{eq:ff1}
\end{equation}
where
\begin{equation}
\Delta_{B_c}=\frac{\sqrt{(s_0^{B_c}-q^2-m_D^2)^2+4m_D^2(m_b^2-q^2)}-(s_0^{B_c}-q^2-m_D^2)}{2m_D^2},\label{eq:delta1}
\end{equation}
and $p^2=m_D^2$ has been used.

The LCSR for the form factor of $B\to Dl\tilde\nu$ has been derived
in Ref.\cite{BD}, here we just give the result:
\begin{eqnarray}
{\cal F}_{B\to
D}(v\cdot v')&=&\frac{2m_b^2}{(m_B+m_D)m_B}\sqrt{\frac{m_D}{m_B}}\frac{f_D}{f_B}e^{m_B^2/M^2}\nonumber\\
&&\times\int_{\Delta_B}^1{\frac{du}{u}\exp{\left[-\frac{m_b^2-(1-u)(q^2-um_D^2)}{uM^2}\right]}
\varphi_D(u)},\label{eq:ff2}
\end{eqnarray}
where
\begin{equation}
\Delta_B=\frac{\sqrt{(s^B_0-q^2-m_D^2)^2+4m_D^2(m_b^2-q^2)}-(s^B_0-q^2-m_D^2)}{2m_D^2}.\label{eq:delta2}
\end{equation}

\section{$D$-meson Distribution Amplitude}

~~~Now let's do a discussion on an important nonperturbative
parameter appearing in the LCSRs, the leading twist DA of $D$-meson,
$\varphi_D(x)$. We reexamine the $D$-meson distribution amplitude
since we missed a factor of $\sqrt{2}$ for the decay constant $f_D$
in determining the coefficients of the DA model \cite{BD}.

The $D$ meson is composed of the heavy quark $c$ and the light
anti-quark $\bar q$. The longitudinal momentum distribution should
be asymmetry and the peak of the distribution should be
approximately at $x\simeq m_c/m_D\simeq0.7$. According to the
definition in Eq.(\ref{eq:da}), $\varphi_D(x)$ satisfies the
normalization condition
\begin{equation}
\int^1_0{dx \varphi_D(x)}=1.
\end{equation}

In the pQCD calculations \cite{pQCD}, a simple model (we call model
I) is adopted as
\begin{equation}
\varphi_D^{(I)}(x)=6x(1-x)(1-C_d(1-2x))\label{eq:da1}
\end{equation}
which is based on the expansion of the Gegenbauer polynomials.
Eq.(\ref{eq:da1}) has a free parameter $C_d$ which ranges from $0$
to $1$, and is supposed to approximate $0.7$ in order to get
consistent results with experiments \cite{pQCD}. Thus we simply take
$C_d=0.7$.

On the other hand, it was suggested in \cite{wf} that the light-cone
wave function of the $D$-meson be taken as:
\begin{equation}
\psi_D(x,\mathbf{k}_\perp)=A_D\exp{\left[-b_D^2\left(\frac{\mathbf{k}_\perp^2+m_c^2}{x}
+\frac{\mathbf{k}_\perp^2+m_d^2}{1-x}\right)\right]}\label{eq:wf}
\end{equation}
which is derived from the Brosky-Huang-Lepage(BHL) prescription
\cite{BHL}. One constraint on the wave function is from the leptonic
decay process $D\to \mu\nu$:
\begin{equation}
\int_0^1\int{\frac{d^2\mathbf{k}_\perp}{16\pi^3}\psi_D(x,\mathbf{k}_\perp)}=f_D/2\sqrt{6}\label{eq:constraint1}.
\end{equation}
Here the conventional definition of the decay constant $f_D$ has
been used, so Eq.(\ref{eq:constraint1}) differs from that in
Ref.\cite{wf} by a factor of $\sqrt{2}$. Another constraint comes
from an estimation of the probability of finding the $|q\bar{q}>$
Fock state in the $D$ meson:
\begin{equation}
P_D=\int^1_0{dx\int{\frac{d^2\mathbf{k}_\perp}{16\pi^3}|\psi_D(x,\mathbf{k}_{\perp})|^2}}\label{eq:constraint2}.
\end{equation}
As discussed in Ref.\cite{wf}, $P_D\approx0.8$ is a good
approximation for the $D$ meson. Based on these two constraints, the
parameters $A_D$ and $b_D^2$ can be fixed. Taking $P_D\approx{0.8}$,
$f_D=222.6 \mbox{MeV}$, $m_c=1.3 \mbox{GeV}$ and
$m_d=0.35\mbox{GeV}$, we have $A_D=225\mbox{GeV}^{-1}$,
$b_D^2=0.580\mbox{GeV}^{-2}$. $\psi_D(x,\mathbf{k}_\perp)$ can be
related to the normalized DA $\varphi_D(x)$ by the definition:
\begin{equation}
\varphi_D(x)=\frac{2\sqrt{6}}{f_D}\int{\frac{d^2\mathbf{k}_\perp}{16\pi^3}
\psi_D(x,\mathbf{k}_\perp)}\label{eq:wf-da}.
\end{equation}
 Substituting Eq.(\ref{eq:wf}) into Eq.(\ref{eq:wf-da}), we
have a model of the DA(model {I}{I})
\begin{equation}
\varphi_D^{({I}{I})}(x)=\frac{\sqrt{6}A_D}{8\pi^2~f_D~b_D^2}x(1-x)\exp{\left[-b_D^2\frac{x
m_d^2 +(1-x)m_c^2}{x(1-x)}\right]},\label{eq:da2}
\end{equation}

Furthermore, as argued in Ref.\cite{pionwf}, a more complete form of
the light-cone wave function should include the Melosh rotation
effect in spin space:
\begin{equation}
\psi^f_D(x,\mathbf{k}_\perp)=
\chi_D(x,\mathbf{k}_\perp)A^f_D\exp{\left[-{b^f_D}^2\left(\frac{\mathbf{k}_\perp^2+m_c^2}{x}
+\frac{\mathbf{k}_\perp^2+m_d^2}{1-x}\right)\right]}
\end{equation}
with the Melosh factor,
\begin{equation}
\chi_D(x,\mathbf{k}_\perp)=\frac{(1-x)m_c+xm_d}{\sqrt{\mathbf{k}^2_\perp+((1-x)m_c+xm_d)^2}}\label{eq:melosh}.
\end{equation}
It can be seen from Eq.(\ref{eq:melosh}) that
$\chi_D(x,\mathbf{k}_\perp)\to 1$ as $m_c\to\infty$, since there is
no spin interaction between the two quarks in the heavy-flavor
meson, ie., the spin of the heavy constituent decouples from the
gluon field, in the heavy quark limit \cite{IS}. However the
$c$-quark is not heavy enough to neglect the Melosh factor. After
integration over $\mathbf{k}_\perp$ the full form of $D$ meson DA
can be achieved (model {I}{I}{I}):
\begin{equation}
\varphi_D^{({I}{I}{I})}(x)=\frac{A^f_D\sqrt{6x(1-x)}}{8\pi^{3/2}f_Db^f_D}y\left[1-Erf\left(\frac{b^f_Dy}{\sqrt{x(1-x)}}\right)\right]
\exp{\left[-{b^f_D}^2\frac{(xm_d^2+(1-x)m_c^2-y^2)}{x(1-x)}\right]}\label{eq:da3},
\end{equation}
where $y=xm_d+(1-x)m_c$ and the error function $Erf(x)$ is defined
as $Erf(x)=\frac{2}{\pi}\int^x_0{\exp({-t^2})dt}$. Using the same
constraints as in Eq.(\ref{eq:constraint1}) and
(\ref{eq:constraint2}), the parameters $A^f_D$ and $b^f_D$ are fixed
as $A^f_D=209\mbox{GeV}^{-1}$ and ${b^f_D}^2=0.540\mbox{GeV}^{-2}$.

\begin{figure}[p]
$$\epsfxsize=0.60\textwidth\epsffile{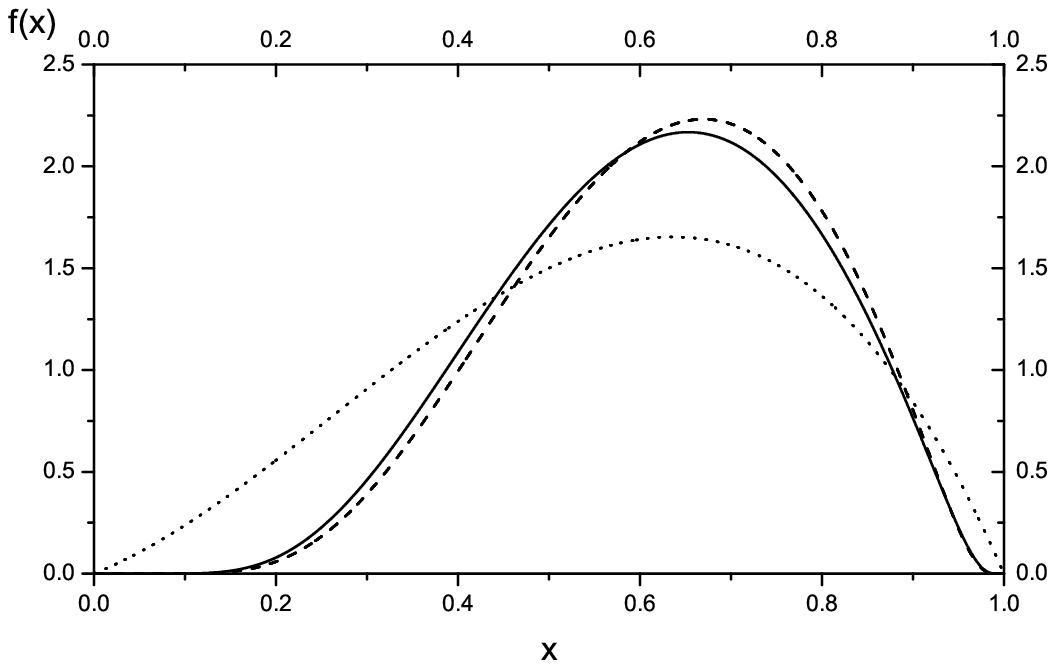}$$
\caption[]{Different kinds of $D$-meson DA's,solid and dashed curves
correspond to model {I}{I}{I} and {I}{I}, while the dotted line
expresses model I.}\label{fig:DA}
\end{figure}

In this paper we will employ the above three kinds of models to do
numerical calculation. All these DA's of the $D$-meson are plotted
in Fig.(\ref{fig:DA}) for a comparison. It can be seen that although
they all have a maximum at $x\simeq0.65$, the shape of them are
rather different, especial in the region $0<x<0.3$ and $0.5<x<0.8$.

\section{Numerical Results and Discussion}

~~~~~~Apart from the DA of the $D$ meson, the decay constant of
$B_c$-meson $f_{B_c}$ is among the important nonperturbative inputs.
For consistency, we use the following corrector
\begin{equation}
K(q^2)=i\int{d^4xe^{iqx}<0|\bar{c}(x)(1+\gamma_5)b(x),\bar{b}(0)(1-\gamma_5)c(0)|0>},
\end{equation}
to recalculate it in the two-point sum rules. The calculation should
be limited to leading order in QCD, since the QCD radiative
corrections to the sum rule for $f_{B_c\to D}(q^2)$ are not taken
into account. The value of the threshold parameter $s^{B_c}_0$ is
determined by a best fit requirement in the region $8
\mbox{GeV}^2{\leq}M^2{\leq}12\mbox{GeV}^2$, where $M^2$ is the
corresponding Borel parameter. The same procedure is performed for
$f_{B}$, in almost the same Borel "window". The results are listed
in Tab.\ref{tab:s0}. As we have ignored all the radiation
corrections, we don't expect our values of $f_{B_c}$ and $f_{B}$ to
be good predictions of that quantity. We use the same threshold
parameters for the corresponding form factors in the LCSRs, except
for the Borel parameter $M_{LC}^2$, which is taken as
$M_{LC}^2\simeq M^2/{<}u{>}$, with ${<}u{>}$ been the average
momentum faraction involved. It turns out that the form factors
depend little on $M_{LC}^2$ in the region $15<M^2_{LC}<20$. The
other input parameters are taken as $m_B= 5.279\mbox{GeV},
m_D=1.869\mbox{GeV},m_{B_c}=6.286\mbox{GeV}$.

\begin{table}

\caption[]{Parameter sets for $f_{B_c}$ and $f_B$, $s^{B_c}_0$ and
$s^B_0$ for $f_{B_c}$ and $f_B$ respectively; $m_b$, $f_{B_c}$ and
$f_B$ are given in GeV, $s^{B_c}_0$ and $s^B_0$ in
$\mbox{GeV}^2$.}\label{tab:s0}
\begin{center}
\begin{tabular}{|l|l|l|l|l|l|}
\hline
      & $m_b$ & $s^{B_c}_0$ &$f_{B_c}$& $s^B_0$ & $f_B$   \\\hline
set 1 & 4.6   &  43.0       &  0.243  & 30.7    & 0.145   \\
set 2 & 4.7   &  42.0       &  0.189  & 30.2    & 0.117   \\
set 3 & 4.8   &  41.2       &  0.137  & 29.8    & 0.090   \\\hline
\end{tabular}
\end{center}
\end{table}

\begin{figure}[p]
$$\epsfxsize=0.60\textwidth\epsffile{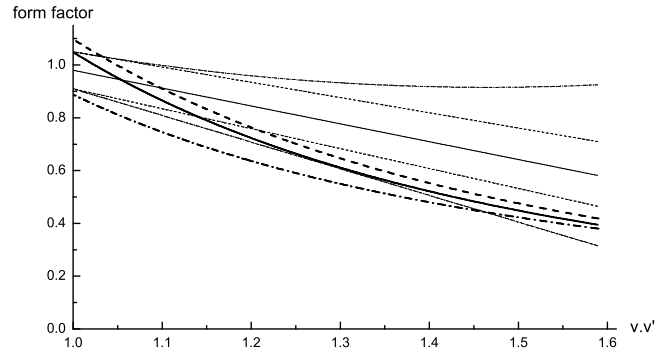}$$
\caption[]{${\cal F}_{B\to D}$ as a function of the velocity
transfer (with the parameters in the set 2). The thin lines
expresses the experiment fits results, the solid line represents the
central values, the dashed(dash-dotted) lines give the bounds from
the linear(quadratic) fits. The thick lines correspond to our
results, with the solid, dashed and dash-dotted lines for model
{I}{I}{I}, {I}{I} and I respectively.}\label{fig:ff1}
\end{figure}

\begin{figure}[p]
$$\epsfxsize=0.60\textwidth\epsffile{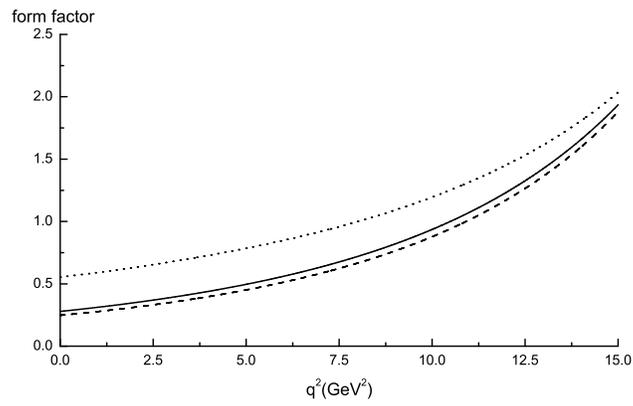}$$
\caption[]{$f_{B_c\to D}(q^2)$ calculated by using different kinds
of DA models. The solid, dashed and dash-dotted lines correspond to
Model III, II and I respectively. Here the threshold parameter set 2
 has been used.}\label{fig:ff2}
\end{figure}

With these inputs, we can carry out the numerical analysis. In
particular, we redo the previous calculation for $B\to Dl\tilde\nu$
in Ref.\cite{BD} and show the corresponding form factor ${\cal
F}_{B\to D}(v\cdot v')$ in Fig.(\ref{fig:ff1}). The result for the
form factor of $B_c \to Dl\tilde\nu$ is given in
Fig.(\ref{fig:ff2}). For ${\cal F}_{B\to D}(v\cdot v')$, similar
results can be obtained by applying the various model DA's at large
recoil region $v\cdot v'\simeq1.59$, ie., $q^2\simeq0$, but rather
different values at the zero recoil point $q^2=q^2_{max}$. It can be
understandable easily from the involved region of the DA. While
$q^2=0$ corresponds to $\Delta_B\simeq0.75$ according to
Eq.(\ref{eq:delta2}), $q^2=q^2_{max}$ corresponds to
$\Delta_B\simeq0.6$, and the models of the $D$-meson DA in the
region $0.5<x<0.8$ are rather different. However, the LCSR result at
the zero recoil point ($q^2=q^2_{max}$) is less reliable, we cannot
get a final conclusion from the difference of the form factor at
this point. Fortunately, the case for $B_c \to Dl\tilde\nu$ is just
opposite, which can be seen from Fig.(\ref{fig:ff2}). There is a big
difference of the form factor at the point $q^2=0$. A detailed
comparison for the form factor at this point with other approaches
is shown in Tab.(\ref{tab:ff0}). The big difference between model I
and others comes from the different contributions of the DA's in the
involved region $0<x<0.45$. Due to the exponential suppression at
the end points, the results from model {I}{I} and {I}{I}{I} are much
smaller than that from model I, and are consistent with the 3PSR
results with the Coulumb corrections included, and the PM result. It
can also be seen from Fig.(\ref{fig:ff1}) and Fig.(\ref{fig:ff2})
that, in both cases, model {I}{I} and {I}{I}{I} actually differ
little, which means that the influence of the Melosh factor is not
so important due to the heavy $c$ quark.

\begin{table}
\caption[]{Form factor $f_{B_c \to D}(0)$ of $B_c\to Dl\tilde\nu$
calculated with different kinds of $D$-meson DA's, in comparison
with that of the 3-Points Sum Rule (3PSR) without \cite{3PSR1} and
with \cite{3PSR2} the Coulumb corrections and Potential Model (PM)
\cite{3PSR2} }\label{tab:ff0}

\begin{center}
\begin{tabular}{l|cccccc}
\hline
                    & medel I & model {I}{I}  &  model {I}{I}{I} &3PSR \cite{3PSR1}& 3PSR \cite{3PSR2} & PM \cite{3PSR2}\\\hline
     $f_{B_c\to D}(0)$  & 0.55   & 0.25  &  0.28  &   $0.13\pm0.05$ &  0.32    & 0.29
     \\\hline
\end{tabular}
\end{center}

\end{table}

The calculated form factor for $B_c\to Dl\tilde\nu$ can be fitted
excellently in the calculated region $0<q^2<15\mbox{GeV}^2$ by the
parametrization:
\begin{equation}
f_{B_c \to
D}(q^2)=\frac{f(0)}{1-a_fq^2/m_{B_c}^2+b_f(q^2/m_{B_c}^2)^2}\label{eq:para}.
\end{equation}
The values of $f_{B_c \to D}(0)$, $a_f$ and $b_f$ are listed in
Table \ref{tab:fit}.

\begin{table}
\caption[]{Form factor $f_{B_c \to D}(q^2)$  in a three-parameter
fit (\ref{eq:para}). The three rows correspond to the calculated
form factors using different sets of parameters, respectively.
}\label{tab:fit}

\begin{center}
\begin{tabular}{lccc}
\hline
            & $f(0)$  &$a_f$  & $b_f$  \\\hline
      set 1  & 0.288   & 3.79  &  4.23  \\
      set 2  & 0.283   & 3.92  &  4.47  \\
      set 3  & 0.288   & 4.03  &  4.77  \\\hline
\end{tabular}
\end{center}

\end{table}

Extrapolating the form factor to the whole kinetic region
$0<q^2<(m_{B_c}-m_D)^2\approx19.5\mbox{GeV}^2$ using this
parametrization, we get:
\begin{equation}
\Gamma(B_c \to Dl\tilde\nu)=(0.197\pm0.013)\times10^{-15}\mbox{GeV},
\end{equation}
and
\begin{equation}
BR(B_c\to Dl\tilde\nu)=(1.35\pm0.05)\times10^{-4}.
\end{equation}
where $\tau(B_c)=0.45\mbox{ps}$ and $V_{ub}=0.0037$ have been used.
The central values are calculated by using the parameters set 2,
while the upper and lower bounds are given by using set 3 and set 1
respectively. Our result for the branching ratio is much larger than
$BR(B_c\to Dl\tilde\nu)=0.4\times10^{-4}$ from Ref.\cite{3PSR2},
they employed a simple pole approximation to extrapolate the form
factor to the whole region. It is also much larger that the PM
result $BR(B_c\to Dl\tilde\nu)=0.35\times10^{-4}$ \cite{PM}, and the
result $BR(B_c\to Dl\tilde\nu)=0.6\times10^{-4}$ from Ref.\cite{CY}.

\section{Summary}

~~~The $B_c$ meson has been observed by the CDF and D0 groups in the
different channels. In this paper we study the weak form factor of
the decay process $B_c(B){\to}D l\tilde{\nu}$ by using the chiral
current correlator within the framework of the QCD light-cone sum
rules, which is similar to the approach for the weak form factor
$f_{B\pi}(q^2)$ in Ref.\cite{cLCSR}. The calculated form factors
depend on the distribution amplitude of the $D$ meson, and we employ
the three different models for the $D$ meson. It has been shown that
the results using the model with a exponential suppression at the
end points are consistent with other approaches. Our results can
also confirm the including of the Coulumb corrections in the 3PSR
calculations for the semileptonic decay $B_c \to Dl\tilde\nu$. In
the LCSRs for the form factors of $B_c(B){\to}D l\tilde{\nu}$, the
involved region of the $D$ meson distribution amplitude is rather
different. Combining the information in the two process, we can find
which model is more suitable for describing the $D$ meson.

We have made a parametrization (\ref{eq:para}) to the form factor by
fitting our calculation in the region $0<q^2<15\mbox{GeV}^2$ , and
the decay width and the branching ratio of the process $B_c{\to}D
l\tilde{\nu}$ have been calculated. It has been shown that
$\Gamma(B_c \to
Dl\tilde\nu)=(0.197\pm0.013)\times10^{-15}\mbox{GeV}$ and $BR(B_c\to
Dl\tilde\nu)=(1.35\pm0.05)\times10^{-4}$. The results are different
from other approaches. It will be expected to test the different
predictions in the coming LHC experiments.

\end{document}